\documentclass[vecphys]{svmult}

\usepackage{makeidx}         
\usepackage{graphicx}        
\usepackage{multicol}        
\usepackage[bottom]{footmisc}
\def\deg{$^{\circ}$}
\def\Msun{M$_{\odot}\,$}
\def\solmyr{M$_{\odot}$yr$^{-1}$}
\def\kms{km s$^{-1}$}

\makeindex             

\begin{document}

\title*{Nuclear spirals in galaxies}
\author{Witold Maciejewski}
\institute{Astrophysics, Denys Wilkinson Bldg, Keble Rd, 
Oxford OX1 3RH, UK \texttt{witold@astro.ox.ac.uk}}
%
%
\maketitle

\begin{abstract}
Recent high-resolution observations indicate that nuclear spirals are often 
present in the innermost few hundred parsecs of disc galaxies. My models show
that nuclear spirals form naturally as a gas response to non-axisymmetry in 
the gravitational potential. Some nuclear spirals take the form of spiral 
shocks, resulting in streaming motions in the gas, and in inflow comparable 
to the accretion rates needed to power local Active Galactic Nuclei. Recently
streaming motions of amplitude expected from the models have been observed
in nuclear spirals, confirming the role of nuclear spirals in feeding
of the central massive black holes.
\end{abstract}

\section{Introduction}
There is an intricate mutual dynamical dependence between the central Massive 
Black Hole (MBH) and the nuclear region of the host galaxy. Centres of galaxies
act as resonant cavities, and the mass of the central MBH contributes to the 
formation and locations of the resonances \cite{m03},\cite{m4a}.
Resonances can either halt or enhance radial gas flow, and thus can control 
fueling of the nucleus \cite{m4b}. Among resonantly-induced features,
nuclear spirals form naturally as a gas response to non-axisymmetry in the 
gravitational potential of a galaxy \cite{e+s},\cite{m4b}. In fact, recent 
high-resolution observations often find in the innermost few hundred parsecs of
disc galaxies nuclear spirals \cite{p+m},\cite{m++} that are likely to continue
inward to within a few parsecs from galaxy's centre \cite{pmr},\cite{f++}.
Because of their ubiquity, nuclear spirals were invoked as the mechanism by 
which material is transported to the central MBH \cite{r+m}. In this paper, I 
present implications of \cite{m4a} and \cite{m4b} for the diagnostic role
of nuclear spirals, and I confront model predictions with the most recent 
observations.

\section{Geometry: indicator of central mass concentration}
Nuclear spirals form naturally as morphology of waves in gas, generated by
a rotating asymmetry in the galactic gravitational potential. Generation 
and propagation of waves is governed by dynamical resonances, whose presence
and positions depend on the central mass distribution in the galaxy. Simple 
linear approximation enables to describe the morphology of nuclear spiral
for most typical rotation curves \cite{m4a}:\\
-- $A$, with a linear inner rise, reflecting solid-body rotation in the 
innermost parts of the galaxy, i.e. constant-density core,\\
-- $B$, same as $A$ but with a central MBH of mass consistent with the 
observed correlations (e.g. \cite{tre}),\\
-- $C$, a pure power-law, corresponding to a central density cusp.\\
These three representative rotation curves are presented in the upper panels 
of Fig.1, while the shapes of the nuclear spirals 
generated by a rotating bisymmetry in the potential are shown in the lower 
panels of Fig.1.

\begin{figure}
\centering
\vspace{-6mm}
\includegraphics[width=\linewidth]{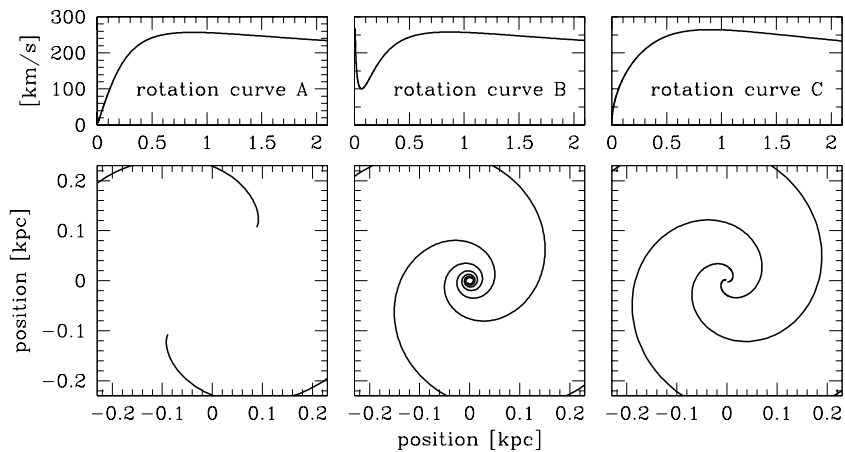}
\vspace{-57mm}
\caption{Rotation curves (upper row), and the corresponding shapes of nuclear 
spiral (lower row), calculated for isothermal gas with 20 \kms sound speed 
for three rotation curves described in the text.}
\label{fig1}       
\end{figure}

If there is no central MBH, and if the rotation curve rises linearly in its 
innermost parts (case $A$), the nuclear spiral will not extend to the centre 
of the galaxy (Fig.1, left column). If there is a MBH in the centre (case $B$),
the nuclear spiral extends to the centre, and it tightly winds around the 
central MBH (Fig.1, central column). If there is a central density cusp (case 
$C$), the nuclear spiral extends to the centre, but it unwinds towards the 
centre (Fig.1, right column). Rotation curves $A$ and $C$ appear similar, but 
their corresponding nuclear spirals are diametrically different. Ubiquity of
nuclear spirals in galaxies may indicate that approximating the inner rise 
of the rotation curve by a straight line is often inadequate. Moreover, nuclear
spiral tightly winding around the galactic centre may indicate the presence 
of a MBH there.

\section{Amplitude: indicator of asymmetries in the potential}
Nuclear spirals are resonant phenomena, and they can be generated by very
small departures from axial symmetry in galaxies. Aside for asymmetries in 
stellar distribution, other asymmetries may contribute to the overall galactic 
gravitational potential. If MBHs in centres of galaxies form by merging of
smaller black holes, then there should be a few black holes of mass one 
or two orders of magnitude smaller than that of the central ones, orbiting 
around the centre of a typical galaxy (e.g. \cite{vhm},\cite{its}).
These black holes constitute a weak perturbation in the gravitational 
potential, which can generate wave phenomena in gas within a disc close to 
the centre of a galaxy. A single orbiting black hole about ten times less 
massive than the central black hole generates a three-arm spiral pattern in 
the central gaseous disc, with density excess in the spiral arms up to 3-12\% 
(\cite{e+m}: see Fig.2, left and central panels). Dusty filaments that have 
been discovered recently in the centres of galaxies 
(e.g. \cite{m++},\cite{pmr}) have luminosity lower from their surroundings 
by $5-10$\%, therefore spiral patterns in gas generated
by the most massive orbiting black holes should be detectable. Interestingly, 
one of the best investigated nuclear
spirals in NGC 1097 (\cite{pmr},\cite{f++}) has three arms
(Fig.2, right panel), difficult to generate by the observed bisymmetric 
bar only.

\begin{figure}
\centering
\vspace{-25mm}
\includegraphics[width=\linewidth]{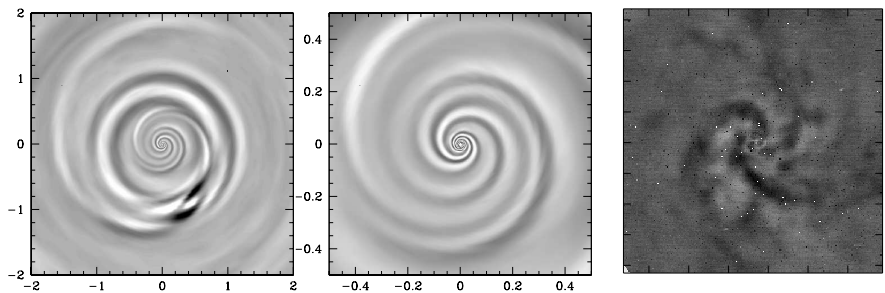}
\vspace{-10.8cm}
\caption{Left and central panels: the density of gas in galactic plane in 
model 11 from \cite{e+m} of a $10^7$\Msun MBH in circular
orbit of 1 kpc radius perpendicular to the galactic plane. Darker shading
represents higher density. Units on axes are in kpc. Right panel: VLT NACO 
J-band image of the nuclear spiral in NGC 1097 after subtraction of radial
intensity gradient by ellipse fitting. The side of the box is 8 arcsec, 
corresponding to about 0.55 kpc.}
\label{fig2}       
\end{figure}

\section{Nuclear spiral as a feeding mechanism of the MBH}
In the present-day Universe, 10-20 per cent of galaxies show nuclear activity 
of Seyfert type. This activity is orders of magnitude weaker than that of 
quasars, and internal, dynamical factors are likely to play a role in 
triggering it. Extensive morphological studies had no success in pointing 
out the mechanism that triggers the nuclear activity, but they might have 
focused on features on too large scales. A typical local Active Galactic
Nucleus (AGN) consumes about 0.01 \Msun of fuel per year (e.g. \cite{pet}),
most likely coming from gas inflow. This corresponds to about $10^6$\Msun 
during its $10^8$-yr 
long activity. Therefore there is no need to transport gas from the outskirts
of a galaxy in order to feed a local AGN, but significant redistribution of 
gas in the innermost tens and hundreds of parsecs should be expected.

\begin{figure}
\centering
\vspace{-4mm}
\includegraphics[width=.78\linewidth]{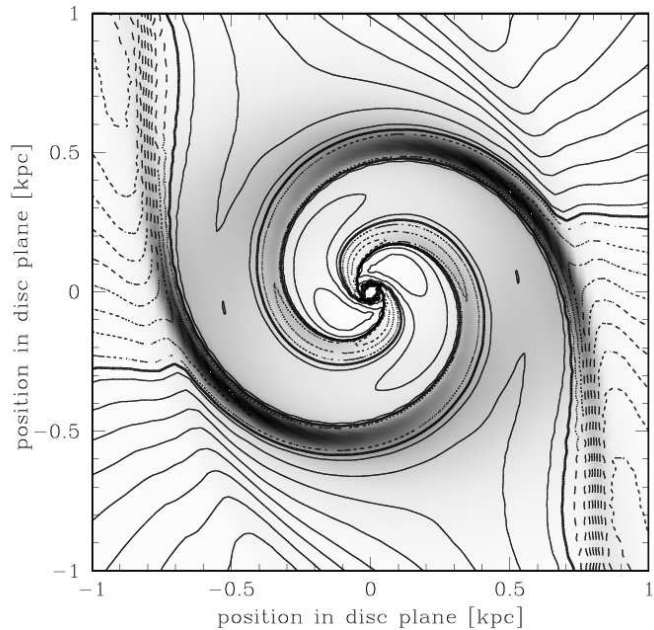}
\vspace{-1mm}
\caption{The density of gas in greyscale in galactic plane in model 8S20 from
\cite{m4b} with nuclear spiral shock. Darker shading represents 
higher density. Contours outline constant radial velocities, and are spaced
every 20 \kms. Outflow (positive radial velocity) is marked by solid contours,
while dashed contours mark inflow (negative radial velocity). Thick solid 
contour marks zero-radial-velocity line.}
\label{fig3}       
\end{figure}

In strong bar, the nuclear spiral has the nature of a shock in gas \cite{m4b}, 
which can trigger gas inflow throughout the spiral. Note 
however, that analogously to gas flow in the region of straight principal 
shocks in the bar (e.g. \cite{a92},\cite{m02}), not all 
gas in the region of nuclear spiral shows radial inflow (Fig.3). In fact, most
of the volume is dominated by outflow of low-density gas, but inflow of dense
post-shock gas in the spiral dominates the budget. Therefore exclusive use
of tracers of dense gas (e.g. molecular emission) can result in biased 
estimates of integrated radial flow of galactic gas.

\begin{figure}
\centering
\vspace{-2mm}
\includegraphics[width=.49\linewidth]{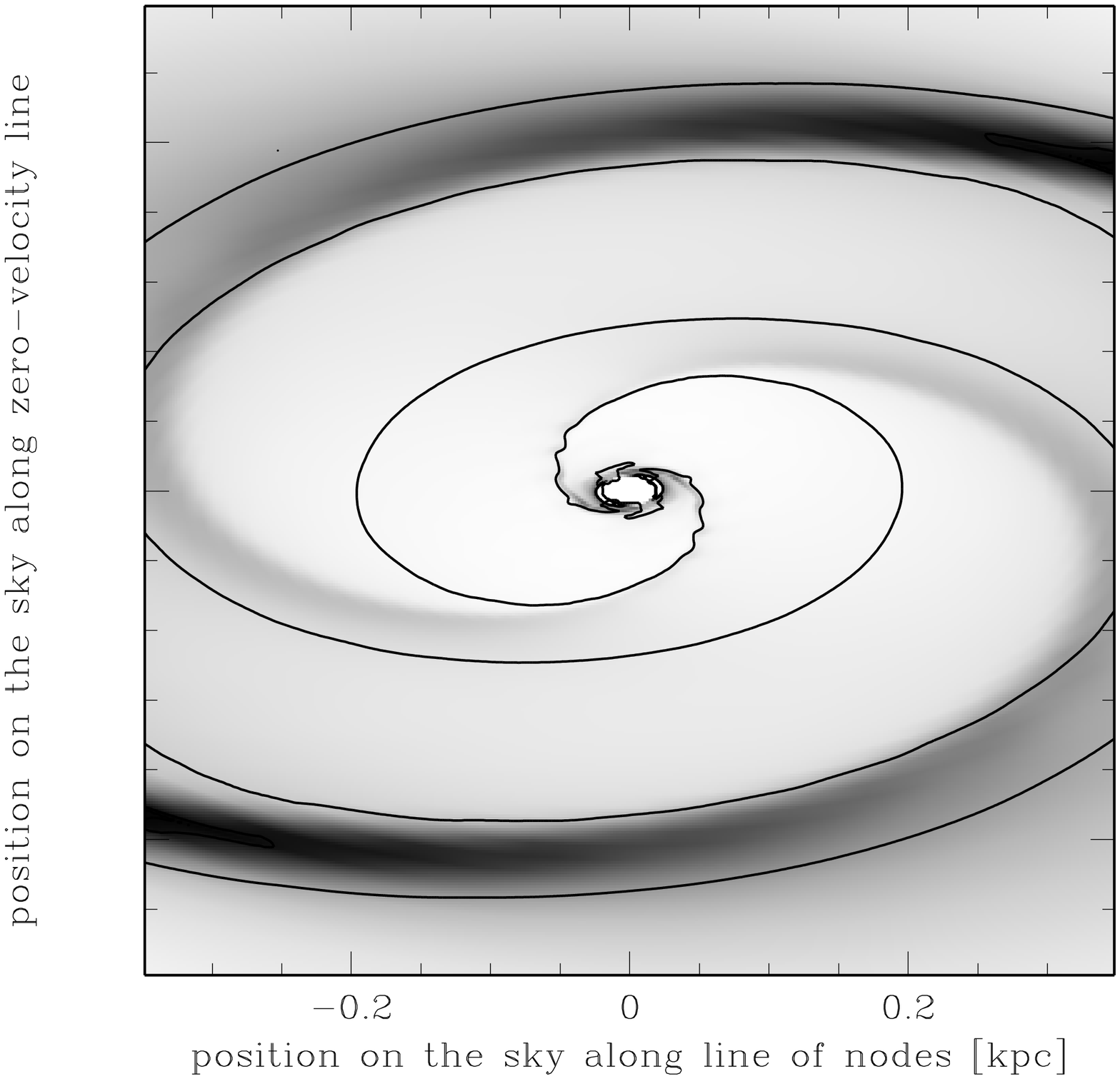}
\includegraphics[width=.49\linewidth]{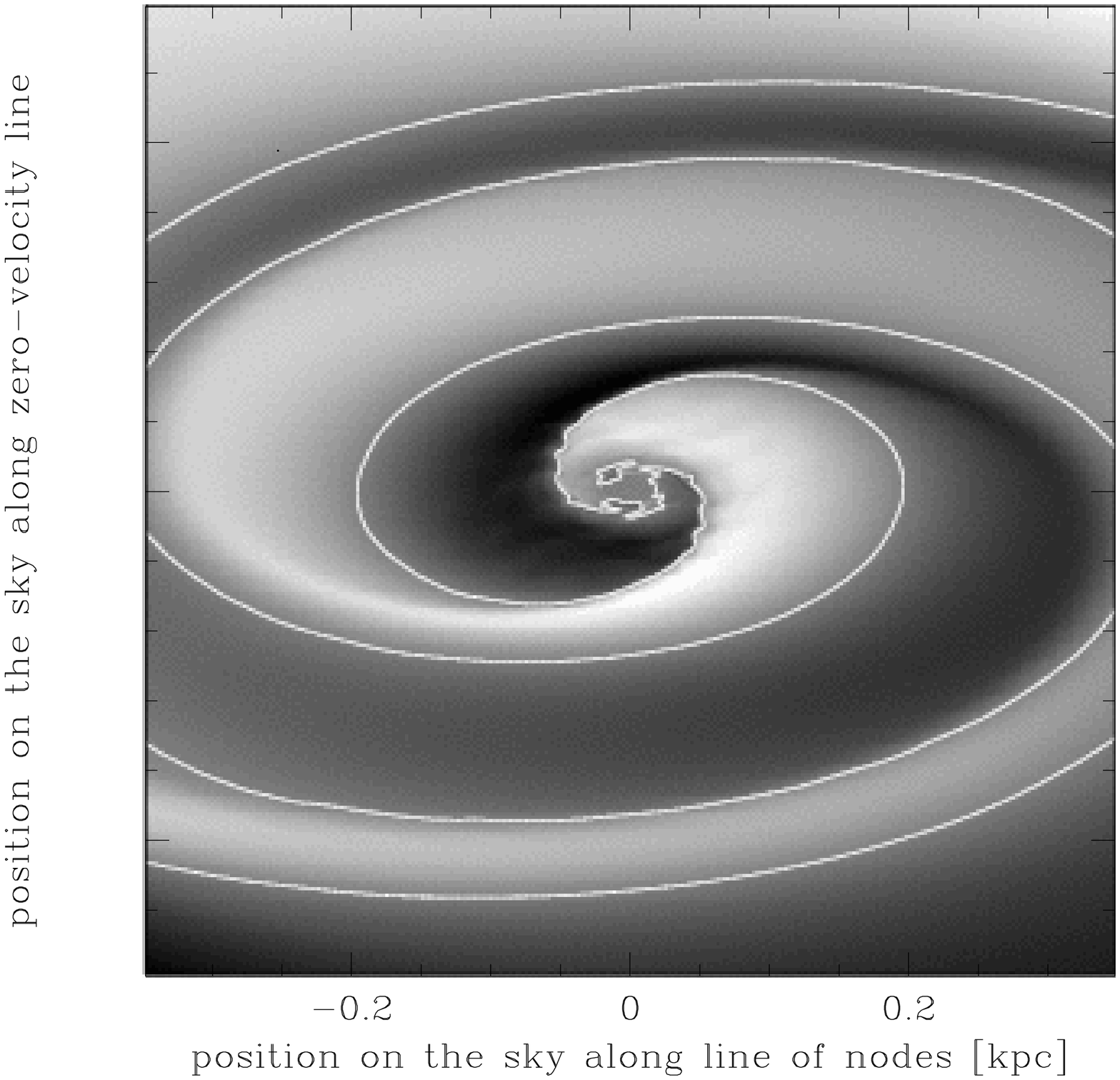}
\vspace{-15mm}
\caption{Left panel: Density in the inner parts of model 8S20 from Fig.3, 
when viewed at inclination of 60\deg. Darker shading represents higher
densities. Right panel: corresponding residual line-of-sight velocities
after subtraction of circular rotation. Darker shading represents negative 
velocities, brighter one -- positive velocities. The solid contours in both 
panels mark zero velocity, and can be used to guide the eye when comparing 
shapes of the spirals.}
\label{fig4}       
\end{figure}

Right panel of Fig.4 shows line-of-sight velocities expected for the standard 
model 8S20 of nuclear spiral from \cite{m4b}, when viewed at 
inclination of 60\deg, and after subtracting contribution of circular motion.
Large streaming motions are expected, of amplitude of up to 50 \kms. This
amplitude can only be achieved when shocks in gas are involved. Recently,
the same algorithm has been applied to the kinematics of nuclear spiral in
NGC 1097 observed with the Integral Field Unit GMOS on the Very Large 
Telescope \cite{f++}. Strong streaming motions of amplitude about
50 \kms\ that show spiral morphology, have been observed. {\it That 
observation confirms presence of shocks in nuclear spirals.} 

Note that the modelled kinematical spiral arms do not overlap with the spiral 
gas morphology (Fig.4, left panel). This is because the matter distribution
is point-symmetric, while the velocity vectors -- point-antisymmetric. The 
shift between the kinematical and morphological spiral arms predicted by
the models is consistent with that observed in NGC 1097 \cite{f++}.

Hydrodynamical models of nuclear spiral shocks indicate the rate of inflow in 
the innermost parsecs of a galaxy up to 0.03 \solmyr \cite{m4b}. This 
inflow is sufficient to feed luminous local AGN, and the feeding can continue 
over long timescales. 
Nuclear spiral shock is less tightly wound than what the linear theory 
predicts, hence loosely wound spirals in the classification developed in 
\cite{m++} may indicate spiral shocks. Interestingly, when in that 
classification
one groups together grand-design nuclear spirals (explicitly linked to shocks
in bars) and loosely wound spirals, they occur considerably more often in
active than in non-active galaxies \cite{m++},\cite{m+2}.

\section{Discussion and conclusions}
Hydrodynamical models of nuclear spirals presented here assume that galactic
gas is a continuous medium, which can be statistically approximated by 
isothermal fluid. One may expect that this approximation breaks down at 
scales small enough, but it is still likely to hold on 10-pc scale, because
continuous dusty nuclear spiral arms extend down to within that distance from 
the nucleus. Perhaps this approximation is most appropriate for the dynamic
interstellar medium, in which dense clouds continuously form and disperse.

The shape of nuclear spiral can serve as an indicator of the presence of a 
MBH in a galaxy's centre, and as an estimator of its mass. The amplitude of 
the spiral can constrain asymmetries in the galaxy's potential, like orbiting 
remnant black holes left from the time of galaxy formation.

Nuclear spirals in galaxies can either be weak density waves which cannot feed 
the nucleus, or strong dissipative shocks, which can generate gas inflow large 
enough to power luminous local AGN. Nuclear spiral shocks should be revealed
in kinematical observations by strong streaming motions, and such motions of
amplitude consistent with theoretical predictions have been recently observed.
This observation makes a sound argument in support of gas inflow in nuclear 
spirals, which can serve as a mechanism feeding the central MBH and leading
the the AGN phenomenon.

This work was partially supported by the Polish Committee for Scientific 
Research as a research project 1 P03D 007 26 in the years 2004--2007.



\printindex
\end{document}